# Unconventional Superconducting Quantum Criticality in Monolayer WTe$_2$

Tiancheng Song[1], Yanyu Jia[1], Guo Yu[1,2], Yue Tang[1], Pengjie Wang[1], Ratnadwip Singha[3], Xin Gui[3], Ayelet J. Uzan[1], Michael Onyszczak[1], Kenji Watanabe[4], Takashi Taniguchi[5], Robert J. Cava[3], Leslie M. Schoop[3], N. P. Ong[1], Sanfeng Wu[1*]

[1]Department of Physics, Princeton University, Princeton, New Jersey 08544, USA.
[2]Department of Electrical and Computer Engineering, Princeton University, Princeton, New Jersey 08544, USA.
[3]Department of Chemistry, Princeton University, Princeton, New Jersey 08544, USA.
[4]Research Center for Functional Materials, National Institute for Materials Science, 1-1 Namiki, Tsukuba 305-0044, Japan.
[5]International Center for Materials Nanoarchitectonics, National Institute for Materials Science, 1-1 Namiki, Tsukuba 305-0044, Japan.

*Correspondence to: sanfengw@princeton.edu

**Abstract**

**The superconductor to insulator or metal transition in two dimensions (2D) provides a valuable platform for studying continuous quantum phase transitions (QPTs) and critical phenomena[1,2]. Distinct theoretical models, including both fermionic and bosonic localization scenarios[1–3], have been developed, but many questions remain unsettled despite decades of research[1,2]. Extending Nernst experiments down to millikelvin temperatures, we uncover anomalous quantum fluctuations and identify an unconventional superconducting quantum critical point (QCP) in a gate-tuned excitonic quantum spin Hall insulator (QSHI)[4–11], the monolayer tungsten ditelluride (WTe$_2$). The observed vortex Nernst effect reveals singular superconducting fluctuations in the resistive normal state induced by magnetic fields or temperature, even well above the transition. Near the doping-induced QCP, the Nernst signal driven by quantum fluctuations is exceptionally large in the millikelvin regime, with a coefficient of ~ 4,100 µV/KT at zero magnetic field, an indication of the proliferation of vortices. Surprisingly, the Nernst signal abruptly disappears when the doping falls below the critical value, in striking conflict with conventional expectations. This series of phenomena, which have no prior analogue, call for careful examinations of the mechanism of the QCP, including the possibility of a continuous QPT between two distinct ordered phases[12–16] in the monolayer. Our experiments open a new avenue for studying unconventional QPTs and quantum critical matter.**

**Main Text**

Monolayer WTe$_2$ is an excellent system for investigating the 2D superconducting QPT in a clean crystalline material with minimal disorder. Distinct from all previous systems studied experimentally, the insulating state of the monolayer is a QSHI[4–6], where recent experiments and theories have shown evidence for an excitonic insulator phase at its charge neutrality point[9–11,17,18]. With electrostatic gating, superconductivity occurs in this 2D crystal above a very low critical carrier density[7,8], on the order of $10^{12}$ cm$^{-2}$. Existing theories suggest that its pairing mechanism could be topological[19] or spin-triplet[20]. The nature of this low-density superconductivity and the intriguing superconductor-to-QSHI quantum transition is currently unknown.

**Vortex Nernst effect in the monolayer**

We examine the superconducting QPT in monolayer WTe$_2$ by measuring the Nernst effect. The Nernst experiment detects a transverse voltage in a material placed in a perpendicular magnetic field ($B$) when a thermal gradient is applied in the longitudinal direction (Fig. 1a). In the case of a type-II superconductor, a magnetic field generates vortices that carry quantized flux and entropy in the superfluid. A temperature gradient ($-\nabla T$) can then drive a flow of vortices, producing phase slippage that engenders a transverse electric field ($E$) due to the Josephson relation. Since its observation in cuprates[21,22] above the critical temperature ($T_c$), the Nernst effect has been widely observed in various superconductors[23]. As a sensitive probe to vortices and superconducting fluctuations, it reveals critical information hidden from electrical transport.

To detect the monolayer Nernst effect, we employ the geometry shown in Fig. 1a. The monolayer WTe$_2$ is fully encapsulated between graphite/hexagonal boron nitride (hBN) stacks to avoid degradation and enable gate-dependent studies (see Methods and Extended Data Fig. 1). The carrier density in the monolayer is modified by gate voltages applied to the top ($V_{tg}$) and bottom ($V_{bg}$) graphite layers. The gate-induced carrier density is $n_g \equiv \varepsilon_r \varepsilon_0 (V_{tg}/d_{tg} + V_{bg}/d_{bg})/e$, where $d_{tg}$ ($d_{bg}$), $\varepsilon_r$, $\varepsilon_0$, and $e$ denotes, respectively, the thickness of top (bottom) hBN dielectric, its relative dielectric constant, vacuum permittivity, and elementary charge. Key to the Nernst experiments are the two microheaters fabricated next to the monolayer, each being a thin narrow metal stripe (~ 200 nm wide, ~ 8 nm thick, and ~ 1 kΩ resistance). We employ a dual-heater measurement scheme[24–27], in which a low-frequency ($\omega$ ~ 13 Hz) alternating current is applied to the two microheaters, with a 90° phase shift between them, to produce an alternating $\nabla T$ with minimal perturbation to the sample temperature ($T$). The Nernst voltage drop ($V_N$) across the two probes is detected at the frequency of 2$\omega$. Unlike in previous thermoelectric measurements on 2D materials where the heater is typically located on the SiO$_2$/Si substrate[28–30], here we fabricate the heaters directly inside the van der Waals stack. This method, which produces a finite $\nabla T$ with minimal heater power ($P_h$) (see Methods and Extended Data Figs. 2-4), enables Nernst measurements at millikelvin temperatures. Our approach can be universally applied to study the thermoelectricity of various 2D crystals and moiré materials.

Figure 1b depicts the electronic phase diagram of the monolayer, based on four-probe resistance ($R_{xx}$) taken on the same device (device 1) (Extended Data Fig. 5). Consistent with previous reports[7,8], it clearly shows both the superconducting and QSHI states with a critical carrier density, $n_{c,R}$ ~ 6.3×10$^{12}$ cm$^{-2}$, at which d$R_{xx}$/d$T$ switches its sign. We characterize this gate-tuned 2D superconductor by the Berezinskii–Kosterlitz–Thouless (BKT) transition temperature ($T_{BKT}$), at which thermal fluctuations unbind vortex-antivortex pairs and produce a finite resistance. For each $n_g$ well above $n_{c,R}$, we extract the corresponding $T_{BKT}$ from its $I$-$V$ characteristics following the usual procedure[31] (Extended Data Fig. 6). $T_{BKT}$, represented by the white dots in Fig. 1b, decreases monotonically with decreasing $n_g$ and vanishes at a density slightly higher than $n_{c,R}$, signifying the $n_g$-tuned QPT.

We first present the Nernst data taken in the highly electron-doped regime ($n_g \sim 2.3 \times 10^{13}$ cm$^{-2}$), where the highest $T_c$ is observed. Figure 1c plots the Nernst signal ($V_N$) together with $R_{xx}$ as a function of $B$, where $V_N$ clearly develops a peak at the field-induced transition to the normal state and vanishes near zero $B$ (i.e., the vortex solid state) as well as at high $B$. With increasing $T$ to high values, $V_N$ disappears (Figs. 1d and e). The Nernst signal, which is antisymmetric with respect to $B$ and coincides with the change in $R_{xx}$, directly probes the motion of vortices in the vortex liquid state[22,23,32]. We further note that the absence of $V_N$ at high $B$ and high $T$ implies that the quasiparticle contributions to the Nernst signal are negligible in our measurements.

**Large quantum fluctuations in the underdoped regime**

Our key results are shown in Fig. 2a, where the Nernst signal is recorded at $T \sim 45$ mK as a function of $n_g$, which is continuously tuned across the superconductor-to-QSHI transition. We discuss important findings near the $n_g$-tuned QPT in three regimes. The first one is the large Nernst signal observed in the underdoped regime at unexpectedly high $B$ (labeled as "I" in Fig. 2a). In Fig. 2b, we present $R_{xx}$ map measured under the same condition, where the black dotted line represents $B_{R,90\%}$, at which $R_{xx}$ drops to 90% of its saturated value at high $B$ (Extended Data Fig. 5). The same dotted line is also plotted in Fig. 2a. Substantial $V_N$ is observed in regime I, well above $B_{R,90\%}$, despite that the $R_{xx}$ data appears to suggest a normal resistive state with no hint of superconductivity. The Nernst data implies that there is a critical field, $B_{c,N}$, above which $V_N$ vanishes (the white dashed line in Fig. 2c). Figures 2c & d highlight the gate dependence of $B_{c,N}$. In contrast to $B_{R,90\%}$, $B_{c,N}$ increases rapidly with decreasing $n_g$ to the critical doping, where $B_{c,N}$, at least $\sim 500$ mT, is more than ten times higher than the corresponding $B_{R,90\%}$. Fig. 2e highlights the heater-power effect on $V_N$. The implications of these observations will be further discussed below.

The prominent Nernst signal in the underdoped regime is enhanced at the lowest $T$. Figure 2f plots the $T$-dependent $V_N$, together with $R_{xx}$, at $n_g \sim 7.0 \times 10^{12}$ cm$^{-2}$. At this doping, the zero-field $R_{xx}$ displays only a slight drop below $\sim 200$ mK but doesn't reach zero. At a $B$ field of 150 mT, this low-$T$ drop is completely suppressed, and an upturn is instead observed. Surprisingly, $V_N$ at this $B$ field remains substantial and survives to temperatures as high as $\sim 700$ mK (Fig. 2f), where traces of superconductivity in $R_{xx}$ vanish at all dopings in this device (Fig. 1b). The Nernst signal above $T_c$ was first observed in cuprates, indicating vortex-like excitations in the pseudogap regime[21,22]. Our observation here occurs in the quantum regime. Figure 2f shows that $V_N$ increases strongly with decreasing $T$. By extrapolation, the Nernst signal is finite in the limit of zero kelvin. It thus arises from quantum fluctuations, instead of thermal fluctuations.

**Giant Nernst effect at the QCP and its sudden death**

Remarkably, Fig. 2a clearly reveals a continuous evolution from the vortex liquid regime at high electron doping to the strongly quantum fluctuating regime near the $n_g$-tuned QPT in a single device. Sharply located at the QPT (labeled as "II" in Fig. 2a), the quantum fluctuations produce a giant Nernst signal near zero $B$, which is another key finding. In Fig. 3a, we carefully examine this regime by plotting $V_N$ under very small $B$. One immediately observes that the Nernst signal is extremely sensitive to the QPT, i.e., a slight detuning of $n_g$ away from $n_{max} \sim 7.2 \times 10^{12}$ cm$^{-2}$, where $V_N$ is maximized, completely suppresses the signal on both sides. To demonstrate that this signal reflects the zero-$B$ property of the QCP, we plot $V_N/B$ in Fig. 3b (only the positive-$B$ side is shown), which may be interpreted as the mobility of the vortices and shows a strong enhancement when $B$ is reduced to near zero. Figure 3c further plots $V_N$ versus $B$ curves at selected $n_g$, in which we confirm that the largest slope occurs at $n_{max}$ near zero $B$. In Methods, we estimate $\nabla T$ to be ~ 5.3 mK/μm, based on which we obtain the corresponding Nernst coefficient, $\nu \equiv (dE/dB)/\nabla T|_{B=0}$, to be ~ 4,100 μV/KT, among the largest values measured in all materials. As a comparison, this giant Nernst coefficient is at least one order of magnitude higher than the typical $\nu$ measured in other superconductors[22,23,32,33]. This large Nernst response near zero $B$ indicates the proliferation of vortices and antivortices when the doping is reduced to the critical value.

The large Nernst signal abruptly disappears right below $n_{c,R}$ ~ 6.3×10$^{12}$ cm$^{-2}$ (labeled as "III" in Fig. 2a, and also in Figs. 3a & b). While the vanishing of $V_N$ above $n_{max}$ is expected in the vortex solid state, its sudden death right below $n_{c,R}$ is a surprise. As a comparison, previous studies in disordered superconducting films have reported substantial Nernst signal deep inside the insulating phase after the transition[33], in clear contrast to the monolayer WTe$_2$. Note that the noisy data that appears below ~ 3.4×10$^{12}$ cm$^{-2}$ in Figs. 2a & 3a is not antisymmetric in $B$ and hence not a Nernst signal (they likely arise due to bad contacts in the more insulating state). The absence of a Nernst signal in regime III is intrinsic, rather than a consequence of resolution limitations in the experiments. This can be seen in Fig. 2e, which reveals a clear contrast in the heater-power dependence of $V_N$ between regimes III and I', and demonstrates that $V_N$ is invariably absent in regime III despite the 10-fold increase in heater power. This is true at all magnetic fields up to 500 mT (see Figs. 2c & d). Regime III is also anomalous in electrical transport. Figure 3d plots $R_{xx}$, where $n_{max}$ and $n_{c,R}$ are both indicated by the red dashed lines. One finds that while $R_{xx}$ displays a dramatic $T$ dependence on the superconducting side, its value in regime III, on the order of ~ kΩ at millikelvin, displays little changes with $T$. This indicates that a metallic-like state resides right below $n_{c,R}$, before the QSHI/excitonic insulator[6,9,10] is fully developed. The sudden death of $V_N$ implies the complete absence of superconducting fluctuations in the metallic-like state directly abutting the QPT. The extreme asymmetry in $V_N$ is our key result.

These highly reproducible findings (see Extended Data Fig. 7 for the data taken on device 2) highlight the intriguing nature of the $n_g$-tuned QPT in monolayer WTe$_2$. To reinforce that the Nernst signal reflects the superconducting quantum fluctuations, in Fig. 4a, we plot the $T$-dependent $V_N$, taken at a fixed small $B$ (2 mT). The amplitude of the signal increases rapidly when $T$ is lowered, suggesting that the Nernst coefficient ($\nu$) attains its maximum at zero kelvin. The corresponding $n_{max}$ shifts its values at different $T$, which traces precisely the $T_{BKT}$ determined from the transport measurement (the same as Fig. 1b). In the canonical phase diagram of a QPT, sketched in the inset of Fig. 4a, the BKT transition traces down to the QCP at absolute zero. The Nernst signal hence directly detects the superconducting QCP in the monolayer. Clearly, quantum fluctuations at the QCP produce the dramatic Nernst signal at the lowest $T$.

**Discussion**

This series of unusual phenomena observed here in monolayer WTe$_2$ has no prior analogue in research on 2D superconducting transitions and challenges the conventional wisdom on QPTs. The standard Landau-Ginzburg-Wilson (LGW) theory[34] describes a continuous transition between an ordered state and a "disordered" state. The order parameter fluctuations are expected to be stronger on the "disordered" side reflecting the suppression of phase fluctuations on the ordered side. In a superconducting transition, a pronounced fluctuation tail penetrating into the normal state is expected and may be further enhanced in the presence of a vortex liquid state above the transition, as widely seen in cuprates[22]. Such conventional expectations of stronger fluctuations on the non-superconducting side (Fig. 4b upper panel) are general, independent of the tuning parameter used for inducing the transition.

In our experiments, we observed three types of superconducting phase transitions in a single device: (a) the $T$-tuned transition; (b) the $B$-tuned transition; and (c) the $n_g$-tuned transition. Indeed, both (a) and (b) follow the above conventional expectations, i.e., pronounced Nernst signals are seen in the normal states (see, e.g., Fig. 2f for the $T$-tuned transition and Figs. 2a & c for the $B$-tuned transition). However, the transition of (c), the $n_g$-tuned QPT, is strikingly opposite to the conventional expectation (Fig. 4b lower panel). The sudden death of the Nernst signal in the normal state below the critical doping $n_c$ (regime III in Fig. 2a) is not anticipated at all in the LGW picture. We conclude that the $n_g$-tuned superconducting QCP in monolayer $WTe_2$ is unconventional.

We further summarize the key experimental features related to this unconventional QCP in Fig. 4c. In the regime $n_g > n_c$, $B_{c,N}$ provides a measure of the pairing strength in the superconducting state. The increase of $B_{c,N}$ implies that the pairing strength increases monotonically as $n_g \rightarrow n_c^+$ (Fig. 4c upper panel), an unexpected behavior. Paradoxically, in spite of the increasing pairing strength, $V_N$ vanishes abruptly once $n_g$ is decreased below $n_c$ (sudden death). One possible way to explain the sudden death is to assume that a new ordered phase emerges in regime III ($n_g < n_c$). This is suggested by the fluctuations in the following two limits. In the limit of $B \rightarrow 0$ and $T \rightarrow 0$ (Fig. 4c middle panel), the giant Nernst response observed at $n_c$ separates two distinct phases with no detectable fluctuations on either side. On the right side, this is understood because the pairing leads to an ordered superconducting state. It is however challenging to explain the absence of fluctuations on the left side if it is treated as a usual metallic-like state. As a comparison, in the limit of large $B$ and $T \rightarrow 0$ (Fig. 4c lower panel), the right side is now in a normal metal state (above $B_{c,N}$). Reducing $n_g \rightarrow n_c^+$, strong Nernst signal occurs and signifies the fluctuations, with a long tail appearing on the right metallic side. This is in sharp contrast to the sudden death of $V_N$ on the left side, even though it displays a similar metallic-like resistivity. The highly asymmetric fluctuations (the lower panel) again seem to suggest a transition from the normal metallic phase (above $n_c$) to an ordered phase (below $n_c$) that does not superconduct. Our experiments, without relying on any model, raise an intriguing question of whether a novel continuous QPT that directly connects two distinct ordered phases is realized here in the monolayer. Note that the metallic phase below $n_c$ (regime III) arises from doping electrons into the excitonic QSHI, which might be an important clue for uncovering its nature.

Examining the possibility of "order-to-order" continuous QPT in a spin model two decades ago has led to the development of the concept of "deconfined QCP" [12,13], which goes beyond the LGW paradigm. Interestingly, the possibility of a deconfined QCP has been explored theoretically at the superconducting transition induced by doping a QSHI[14–16] in a toy model. There is strong interest in searching for systems hosting a QCP beyond the LGW description[12–16,35–40], but their experimental detection is difficult. Independent of theoretical developments, our experimental work calls for a careful examination of the novel QCP discovered here (see Methods for more discussions). Several other 2D systems have also shown interesting quantum criticality[41–44], while the physics is distinct from the case reported here. Our experiments show that, in general, fluctuations in the vicinity of a QCP provide a powerful probe to the nature of a QPT.

**Acknowledgments**

We acknowledge helpful discussions with Z. Bi, T. Grover, F. D. M. Haldane, D. Huse, B. Lian, S. Ryu, S. Sondhi, A. Vishwanath, A. Yazdani, and Y. Zhang. We especially thank L. Pfeiffer and M. Shayegen for their GaAs sample, which was used for calibrating the electron temperature of our dilution refrigerator. This work is mainly supported by ONR through a Young Investigator Award (N00014-21-1-2804) to S.W. Measurement systems and data collection are partially supported by NSF through a CAREER award (DMR-1942942) to S.W. Data analysis is partially supported by AFOSR Young Investigator program (FA9550-23-1-0140) to S.W. N.P.O. is supported by the U.S. Department of Energy (DE-SC0017863). S.W. and L.M.S. acknowledge support from the Eric and Wendy Schmidt Transformative Technology Fund at Princeton. Materials synthesis and device fabrication are partially supported by the Materials Research Science and Engineering Center (MRSEC) program of the National Science Foundation (DMR-2011750) through support to R.J.C., L.M.S., N.P.O., and S.W. T.S. acknowledges support from the Princeton Physics Dicke Fellowship program. A.J.U. acknowledges support from the Rothschild Foundation and the Zuckerman Foundation. K.W. and T.T. acknowledge support from the JSPS KAKENHI (Grant Numbers 19H05790, 20H00354, and 21H05233). L.M.S. and N.P.O. acknowledge support from the Gordon and Betty Moore Foundation through Grants GBMF9064 and GBMF9466, respectively. L.M.S is also supported by the David and Lucile Packard Foundation and the Sloan Foundation.

**Author contributions**

S.W. designed the project. T.S. fabricated the devices and performed measurements, assisted by Y.J., Y.T., G.Y., P.W., A.J.U., and M.O. G.Y. and P.W. built the dilution refrigerator measurement system. T.S., S.W., and N.P.O. analyzed the data. R.S., L.M.S., X.G., and R.J.C. grew and characterized bulk WTe$_2$ crystals. K.W. and T.T. provided hBN crystals. S.W., T.S., and N.P.O. interpreted the results and wrote the paper with input from all authors.

**Competing interests**

The authors declare that they have no competing interests.

**Data availability**

All data needed to evaluate the conclusions in the paper are presented in the paper. Additional data related to this paper are available from the corresponding author upon reasonable request.

Fig. 1

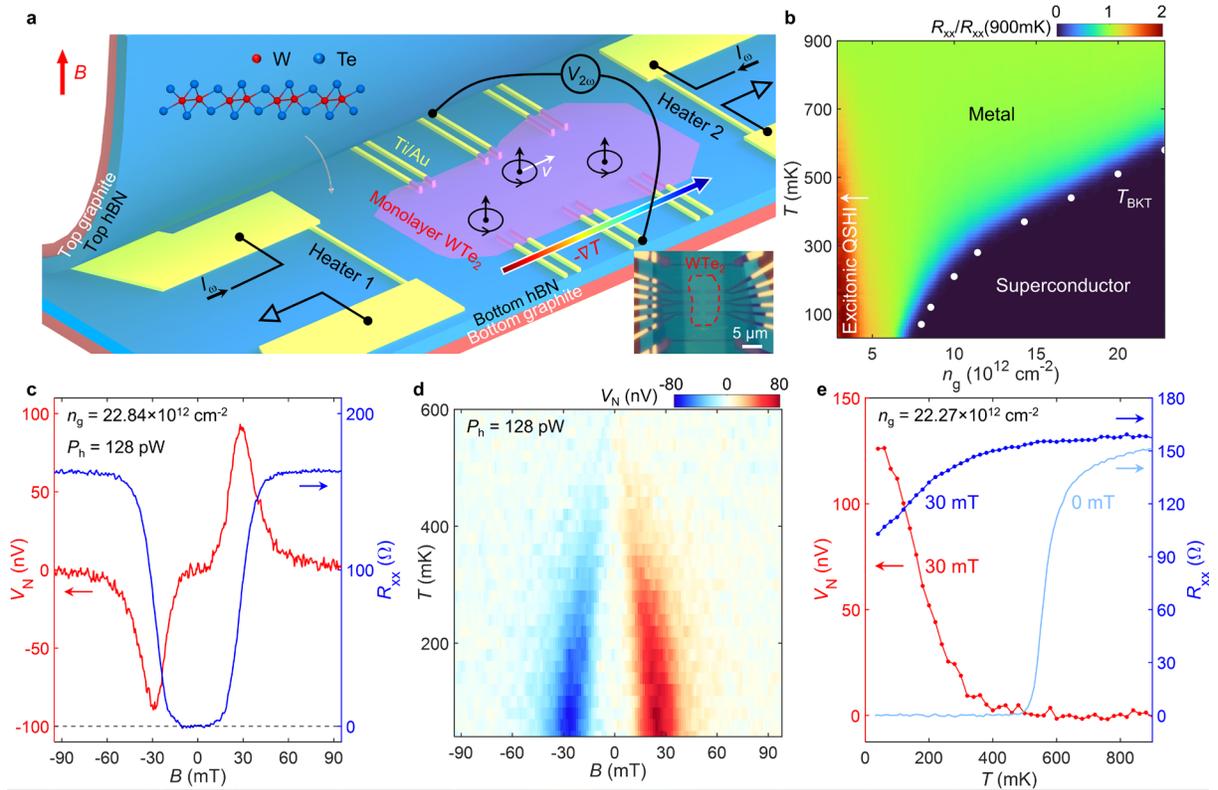

**Fig. 1 | Vortex Nernst effect and electronic phase diagram of monolayer WTe$_2$. a**, Cartoon illustration of the device structure for measuring Nernst signal. Current is applied to the two microheaters to produce a temperature gradient on the monolayer WTe$_2$. The inset shows the optical microscope image of device 1. The monolayer WTe$_2$ flake is outlined in red. **b**, Four-probe resistance as a function of $n_g$ and $T$, measured on the same device. For each $n_g$, $R_{xx}$ is normalized to its value at 900 mK to highlight its temperature dependence (also see Extended Data Fig. 5). The white dots represent $T_{BKT}$. **c**, Nernst signal (red) and $R_{xx}$ (blue) as a function of magnetic field ($B$) for $n_g = 22.84\times10^{12}$ cm$^{-2}$. The dual-heater power ($P_h$) is 128 pW. **d**, $B$-dependence of $V_N$ as a function of $T$ for the same $n_g$. **e**, $T$-dependence of $V_N$ (red) and $R_{xx}$ (blue) measured at 30 mT for $n_g = 22.27\times10^{12}$ cm$^{-2}$. The zero-field $R_{xx}$ is shown as a reference (light blue).

**Fig. 2**

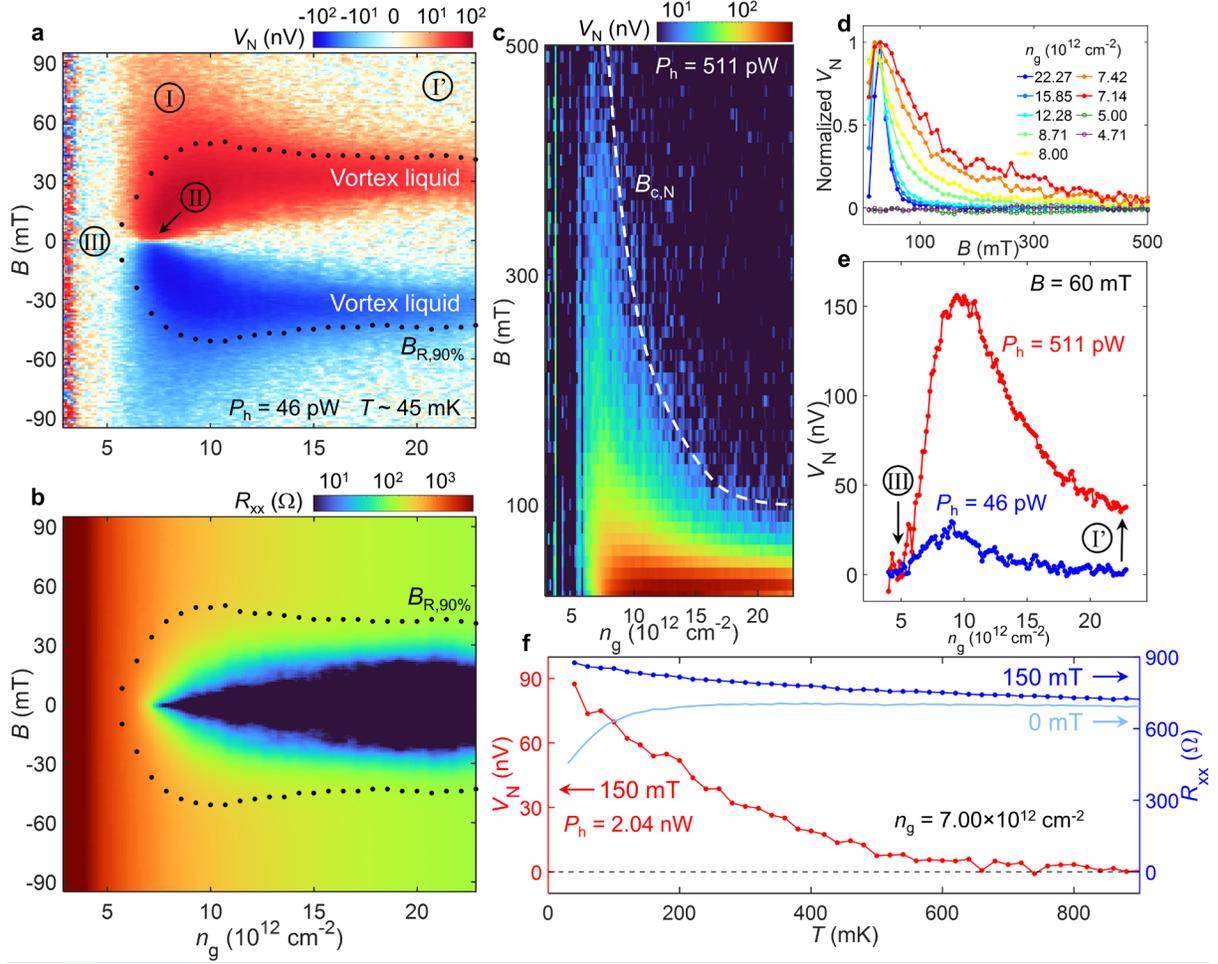

**Fig. 2 | Gate-tuned Nernst effect and superconducting fluctuations well above critical magnetic field and critical temperature. a**, Nernst signal as a function of $n_g$ and $B$. Three different regimes of interest are labeled as I, II, III, respectively. **b**, $R_{xx}$ as a function of $n_g$ and $B$ measured under the same condition as **a**. The black dotted line represents $B_{R,90\%}$, at which $R_{xx}$ drops to 90% of its saturated value at high $B$. **c**, Nernst signal measured up to high $B$, with a higher power ($P_h = 511$ pW). The white dashed line indicates the critical field, $B_{c,N}$, above which $V_N$ vanishes. **d**, $n_g$-dependence of the Nernst signal versus $B$. $V_N$ is normalized to its peak value. **e**, Heater-power effect on the Nernst signal. The data are horizontal line cuts of **a** and **c**, respectively, at $B = 60$ mT. **f**, $T$-dependence of $V_N$ (red) and $R_{xx}$ (blue) measured at 150 mT for a selected low density near the QPT ($n_g = 7.00 \times 10^{12}$ cm$^{-2}$). The zero-field $R_{xx}$ is shown as a reference (light blue).

Fig. 3

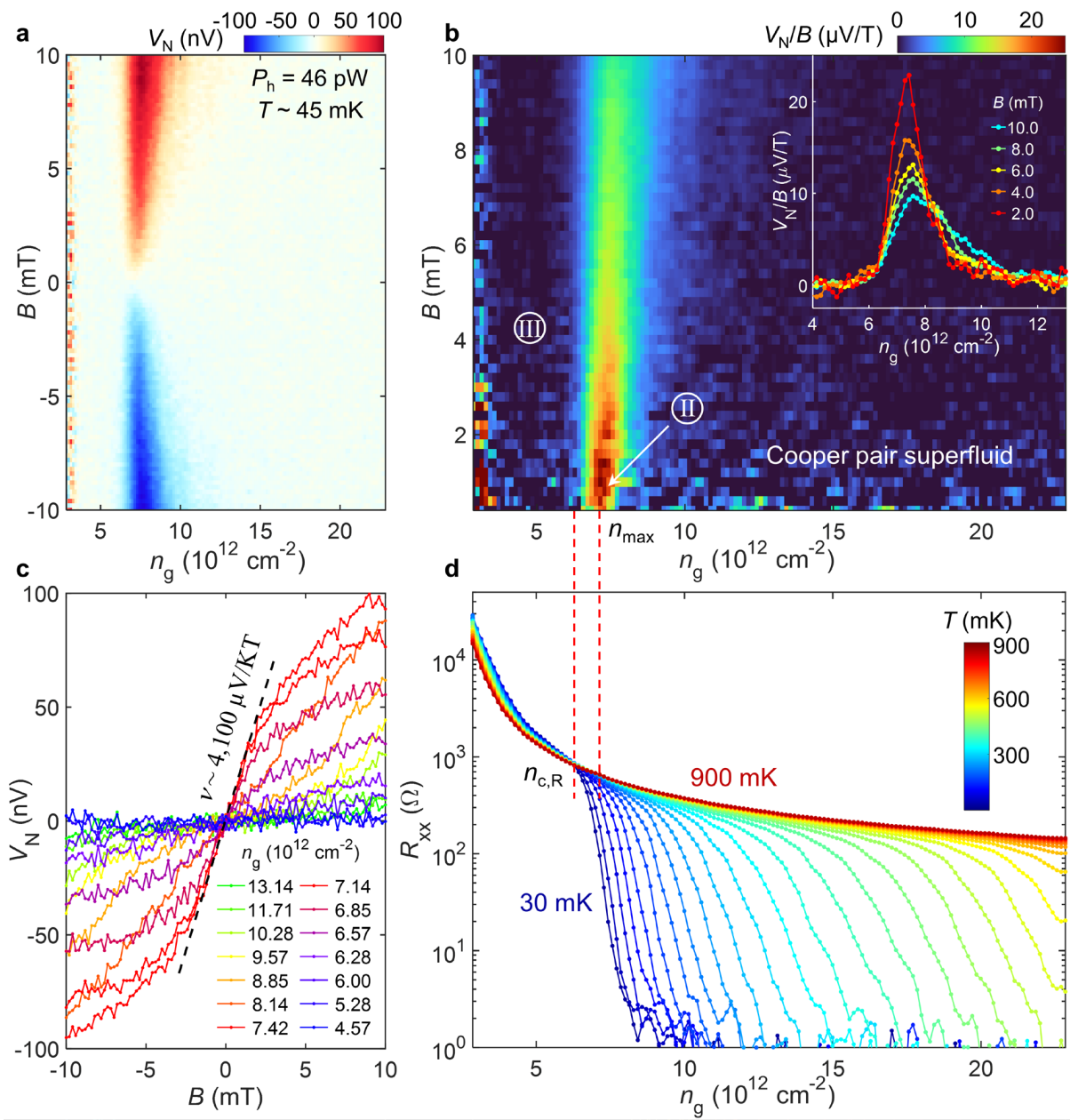

**Fig. 3 | Giant Nernst effect at the QCP and its sudden death. a**, Nernst signal as a function of $n_g$ under very small $B$. **b**, $V_N/B$ as a function of $n_g$ and $B$, calculated from **a**. Only the positive-$B$ side is shown. Inset displays selected line cuts of the same data. **c**, $V_N$ versus $B$ curves at selected $n_g$. The dashed line extracts the largest slope, which occurs at $n_{max}$ near zero $B$. The estimated $\nu$ is indicated. **d**, $T$-dependence of $R_{xx}$ versus $n_g$. The two red dashed lines indicate $n_{c,R}$ and $n_{max}$, respectively.

**Fig. 4**

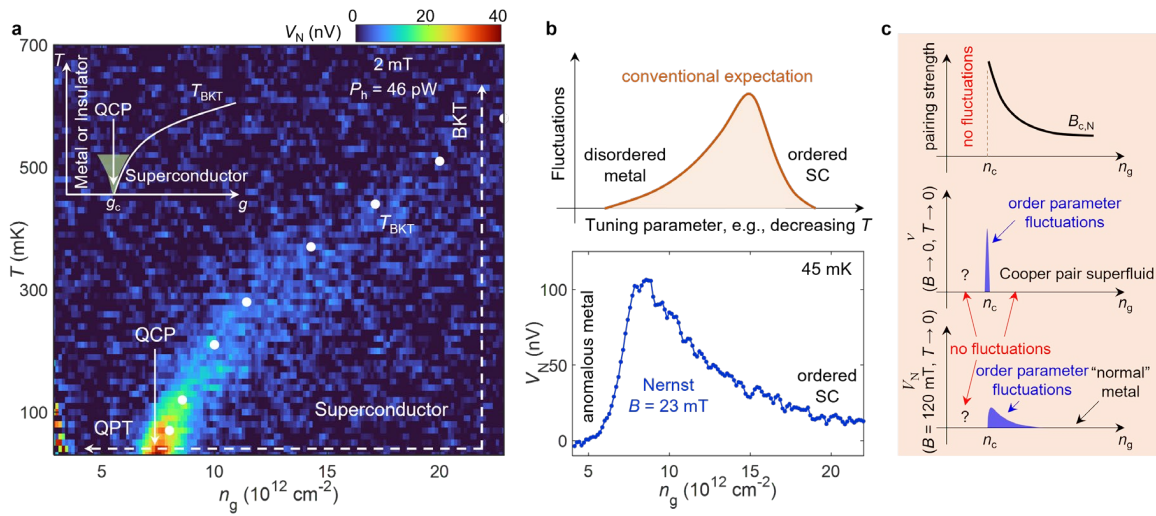

**Fig. 4 | QPT and anomalous quantum fluctuations detected by Nernst effect. a**, Nernst signal as a function of $n_g$ and $T$. The white dots represent $T_{BKT}$. The inset sketches the typical phase diagram of a QPT, where the tuning parameter ($g$) is realized by $n_g$ in our case. **b,** Contrast between the Nernst signal (lower panel, data extracted from Fig. 2a) and the expected fluctuations from the conventional LGW theory (upper panel). **c**, Illustrations of some key findings summarized from Figs. 2 and 3.

# Methods

## Device fabrication

The device fabrication process was separated into two parts: bottom and top, and followed closely to the previous reports[9,10]. (i) Bottom: The hBN and graphite flakes were mechanically exfoliated onto 285 nm $SiO_2$/Si substrates and examined by optical and atomic force microscopy. Only atomically clean and smooth flakes with desirable thickness were used for making devices. The bottom hBN/graphite stacks were made using the standard dry transfer technique and released on $SiO_2$/Si substrates with pre-patterned Ti/Au (5/100 nm) metal pads for wire bonding. After dissolving the polymer in chloroform, Ti/Au (2/6 nm) electrodes and microheaters were deposited on top of the bottom stacks using standard electron beam lithography with a bilayer resist and electron beam evaporation. This was followed by another step of electron beam lithography and electron beam evaporation to deposit Ti/Au (5/60 nm) electrodes to connect the thin electrodes on the bottom stacks to the pre-patterned thick wire-bond metal pads. The surface of the electrodes on the stacks was finally cleaned using the contact mode of an atomic force microscope (AFM). (ii) Top: High-quality $WTe_2$ crystals were exfoliated onto 285 nm $SiO_2$/Si substrates in an inert gas (Argon) glovebox with water and oxygen concentration less than 0.1 ppm. Monolayer $WTe_2$ flakes were identified by optical contrast. The top graphite/hBN stacks were made using the same dry transfer technique. The monolayer $WTe_2$ flakes were then aligned with the top stacks and picked up. The top stacks were finally released on the bottom stacks, and the monolayer $WTe_2$ flakes were fully encapsulated and in contact with the bottom electrodes. The polymer was dissolved in chloroform for less than five minutes to minimize exposure to ambient conditions. The microscope and AFM images during the device 1 fabrication process are shown in Extended Data Fig. 1.

## Electrical transport measurements

The devices were loaded into a dilution refrigerator equipped with a superconducting magnet and a base temperature of ~ 20 mK. Four-probe resistance measurements were performed using the standard AC lock-in technique with a frequency of ~ 13 Hz and an AC current excitation of ~ 5 nA.

## Electron temperature calibration

The measurement wires in the dilution refrigerator are heavily filtered by Thermocoax cables. The electron temperature of the measurement condition was examined and calibrated[45] by the activation behavior of the fragile fractional quantum Hall states in an ultra-high mobility GaAs quantum well, i.e., $R_{xx} \propto \exp(-\Delta/2k_B T_e)$, where $R_{xx}$ is the four-probe longitudinal resistance, $\Delta$ is the energy gap, $k_B$ is the Boltzmann constant, and $T_e$ is the electron temperature. Extended Data Fig. 2 shows $R_{xx}$ in log scale versus the inverse fridge temperature ($1/T$), taken at the 14/9 fractional quantum Hall state. Below 45 mK, the electron temperature of the device deviates from the fridge temperature. We estimated the base electron temperature to be ~ 32 mK at a base fridge temperature of ~ 24 mK. Otherwise mentioned, the temperature mentioned throughout this paper refers to the electronic temperature.

## Nernst measurements

The Nernst measurements were performed on the same devices in the same dilution refrigerator. As illustrated in Fig. 1a, the two microheaters were fabricated next to the monolayer WTe$_2$ flake, each being a thin narrow metal stripe (~ 200 nm wide and ~ 8 nm thick) with a low-temperature resistance of ~ 1 kΩ. We employed a dual-heater measurement scheme, and alternating current was applied to the two microheaters with the same frequency (ω ~ 13 Hz), but a 90° phase shift between each other[24–27]. This produced an alternating temperature gradient (∇T) while minimizing the change of the sample temperature itself near the measurement probes. The Nernst voltage drop ($V_N$) across the two probes is detected at the frequency of 2ω. The dual-heater power was as low as 46 pW in device 1 to generate ∇T and measure $V_N$, thanks to our vdW-heating device geometry. When the two microheaters were on, the sample temperature was expected to be a bit higher. At the base temperature of ~ 20 mK, when the two microheaters are supplied with a power of 46 pW, we estimate the local sample electron temperature of device 1 to be ~ 45 mK using the approach described below.

## Temperature gradient estimation

Here we estimate the temperature gradient on monolayer WTe$_2$ generated by the microheaters. Instead of using external thermometers, we used the local two-probe resistance of monolayer WTe$_2$ itself to measure the in situ local temperature of the flake. This can be achieved because, at low carrier densities (the insulating state), monolayer WTe$_2$ exhibits a strong temperature-dependent resistance, making it a good thermometer.

As shown in Extended Data Fig. 3a, two pairs of probes (1 and 4 were chosen as Pair 1, 2' and 3' were chosen as Pair 2) near the Nernst probes (located in the middle) were selected to measure the local temperatures ($T_1$ and $T_2$). When Heater 1 is on and Heater 2 is off, a temperature difference is generated between Pair 1 and Pair 2, as illustrated in Extended Data Fig. 3b, and $T_1 > T_2$. Similarly, when Heater 1 is off and Heater 2 is on, the temperature gradient is reversed, and $T_1 < T_2$. Here, we show our procedure for estimating the temperature gradient when a DC current ($I_h$) is applied to Heater 2 while Heater 1 is off. Extended Data Figs. 3c and d show the two-probe resistance for Pair 2 and Pair 1, respectively, as a function of the Heater 2 current ($I_h$), at different fridge temperatures ($T$). The black dashed lines show the contour plots which trace the same $R_{2p}$, indicating the same local temperature of each pair. Following the black dashed line, a Heater 2 current can be tracked to a fridge temperature with both heaters off, which implies the local sample temperature near the pair. As a result, for a given Heater 2 power, a $T_{2p}$ can be extracted for Pair 2 and Pair 1, respectively. Clearly, the heater power has a stronger effect on Pair 2 compared to Pair 1, because Heater 2 is closer to Pair 2. As an example, for the 100 pW power (DC current) at a fridge temperature ($T$) ~ 20 mK, the estimated local sample temperatures ($T_1$ and $T_2$) are ~ 120 mK and ~ 60 mK.

For the Nernst experiment, we employed a dual-heater measurement scheme and used the AC lock-in technique. Alternating current was applied simultaneously to the two microheaters with the same low frequency (ω ~ 13 Hz), but a 90° phase shift between each other. The two currents are: $I_1 = I_h \sin(\omega t)$ and $I_2 = I_h \sin(\omega t + \frac{\pi}{2})$, and thus the powers are: $P_1 = \frac{1}{2}I_h^2 R_{h1}(1 - \cos(2\omega t))$ and $P_2 = \frac{1}{2}I_h^2 R_{h2}(1 + \cos(2\omega t))$, where $R_{h1}$ and $R_{h2}$ are the resistances of Heater 1 and Heater 2 (~ 1 kΩ). This corresponds to an alternating temperature gradient, $\nabla T = \frac{\Delta T}{d}\sin(2\omega t - \frac{\pi}{2})$, where $\Delta T = (T_1 - T_2)$ is the extracted temperature difference and $d$ is the distance between the two pairs which is ~ 5.2 μm. The Nernst voltage drop ($V_N$) is detected at the frequency of 2ω with a crest factor ($\sqrt{2}$). The Nernst coefficient is defined as $\nu \equiv (E/B)/\nabla T$, where $E = \frac{V_N}{w}$ and $w$ is the distance between the two probes which is ~ 1.5 μm. For the 46 pW AC dual-heater power, the $\nabla T$ is estimated to be ~ 5.3 mK/μm. For Fig. 3c, the corresponding Nernst coefficient is determined to be ~ 4,100 μV/KT.

### High efficiency of microheaters fabricated on vdW heterostructures

Here, we demonstrate that the location of the microheaters is key to measuring thermoelectricity of 2D crystals in a vdW stack. As shown in Extended Data Fig. 4a, we fabricated device 2 with a similar geometry to device 1 shown in the main text, but with additional microheaters located on and off the vdW stack. For a direct comparison, three microheaters are placed very close to each other, while Heater 1 is fabricated on top of both bottom hBN and graphite, Heater 2 is only in contact with bottom hBN, and Heater 3 is only in contact with $SiO_2$/Si substrate. Extended Data Fig. 4b shows the Nernst signal as a function of $B$ measured using Heater 1 (red), Heater 2 (blue), and Heater 3 (black), respectively, using the same microheater power. One clear finding is that Heater 3 generates the smallest signal that can be barely detected. While Heater 2 generates a good signal, it is significantly lower compared to Heater 1. The data suggest that Heater 1 is most efficient in generating $\nabla T$. This can be understood because at ultralow temperatures, graphite, as a metal, is a good thermal conductor. The idea here is to engineer the thermal path within the vdW stack, while avoiding the poor thermal path through $SiO_2$. We also intentionally chose the long and narrow rectangular graphite flake to enhance the uniformity of $\nabla T$. In the experiments, the application of a heater power also raises the local temperature near the probe, in addition to generating $\nabla T$. Hence to study QPT, the lowest possible heater power is preferred in order to minimize the sample temperature.

In device 1, we employ such a vdW-heating strategy and further optimize its performance by placing the heater very close to the monolayer sample. For device 2, this sample-heater distance is ~ 20 μm, while in device 1 it is only ~ 5 μm. As a result, we are able to generate efficiently a temperature gradient with an ultralow power (~ 46 pW), enabling the Nernst experiment on the monolayer sample down to the millikelvin regime.

### Berezinskii–Kosterlitz–Thouless (BKT) transition temperature

Extended Data Fig. 6a shows the $I$-$V$ curves in device 1 at various temperatures, which exhibit the characteristic nonlinear behaviors for a 2D superconductor. The black dashed line shows a tentative fit to the $V_{xx} \propto I_{DC}^3$ power law[31], which extracts $T_{BKT}$ to be ~ 580 mK for $n_g = 22.84 \times 10^{12}$ cm$^{-2}$. The corresponding differential resistance (d$V_{xx}$/d$I$) as a function of $I_{DC}$ and $T$ is also shown in Extended Data Fig. 6b. Similarly, for various carrier densities, $T_{BKT}$ is extracted by fitting to the $V_{xx} \propto I_{DC}^3$ power law and shown in Extended Data Fig. 6c.

### Discussion of the sudden death syndrome in the Nernst signal

We first emphasize that, at this stage, our aim is to report the striking experimental findings rather than to identify the correct theory. The explanation to our experimental results is widely open at this stage. In the main text, we discuss that the experimental phenomena observed here associated with the $n_g$–induced superconducting QCP has no prior analogue and hence requires an unusual explanation, especially regarding the sudden death of the Nernst signal right below the critical doping, $n_c$. We have argued that the sudden death may be explained if a new "ordered" non-superconducting phase is formed on the side right below $n_c$ (i.e., regime III in Fig. 2a). If it were the usual "disordered" normal phase, fluctuations would be present (e.g., like the normal state induced by $B$, above $B_{R,90\%}$). At this stage, we have no experimental information on what the order might be. Future experiments and theories are necessary to clarify the situation. The experimental data raise the intriguing question on whether an "order-to-order" continuous QPT is realized here.

Examining the same question on a spin model led to the proposal of "deconfined" quantum criticality two decades ago[12,13]. Interestingly, there exists a theoretical proposal of a deconfined QCP (DQCP) at QSHI-to-superconductor transition[14–16], in a toy model that requires an exotic assumption of a special kind of "order" formed spontaneously on the non-superconducting side. Its relevance to our expriments remains to be clarified. Nevertheless, we provide a discussion in the context of DQCP, evaluating both its promising attributes and drawbacks in explaining our experiments.

We summarize some key features of a DQCP[12–16]. (a) It separates two ordered phases (i.e., different spontaneous symmetry-breaking phases); (b) Topological defects on one side (e.g., vortices in the superconducting state) proliferate at the transition, destroy the order, and then condense to form a new ordered phase on the other side. In the toy-model proposal of the doping-induced QSHI-to-superconductor transition[14–16], the topological defects are charged skyrmions on the QSHI side and vortices on the superconductor side; and (c) The phase transition is continuous.

Now we compare our experiments with these key DQCP features. (a) The superconducting state above the critical doping, $n_c$, is surely ordered. We have argued that the sudden death seems to indeed indicate a non-superconducting ordered phase on the other side below $n_c$ (i.e., regime III); (b) Our Nernst experiments directly probe the topological defects on the superconducting side (i.e, the vortices). It is intriguing if one examines what happens to the Nernst signal at the lowest $T$ and lowest $B$ (Figs. 3a & b). In the superconducting state, no Nernst signal is detected (i.e., ordered, no fluctuations). When $n_g$ is decreased towards the QCP from the superconducting side, the Nernst signal increases rapidly, implying the proliferation of vortices. However, once $n_g$ crosses the QCP, the Nernst signal vanishes abruptly. The experiment raises the important question of what happened to the vorticity. The data appear to imply that they transform into a new phase (regime III), which seems to be an ordered one in order to suppress the fluctuations, consistent with a DQCP scenario; (c) A superconductor-to-metal/insulator transition in 2D, such as the one that we are examining, is by default continuous. We elaborate more on this point (c) in the following.

A DQCP is a proposal for a direct continuous transition between two ordered phases, thus does not follow the standard Laudau-Ginzberg-Wilson (LGW) theory; the latter describes a continuous phase transition between an "ordered" phase (with a finite order parameter $\Delta$) and a "disordered" phase ($\Delta = 0$). However, for the same reason, the DQCP[12–16] confronts the challenge of how to distinguish it from other scenarios that connect two ordered phases, in particular (*i*) a first-order transition or (*ii*) two independent transitions that are accidentally close to each other (namely, the transition in fact occurs in two steps, one from the ordered phase 1 to a disordered phase via an LGW transition and then immediately the other transition from the disordered phase to a new ordered phase 2 via another LGW transition). Indeed, all theoretical proposals on DQCP face this challenge even at the purely theoretical level, and debates are ongoing. Ruling out these two scenarios is a key step for identifying a DQCP.

We first note that in our case neither scenario (*i*) nor (*ii*) explains the central puzzle presented in the data, which is the sudden death of the Nernst signal below the critical density $n_c$. In fact both scenarios assume an ordered phase that suppresses superconducting fluctuations below $n_c$ (regime III). Hence scenarios (*i*) and (*ii*) shares the same assumption needed for a DQCP from this point of view. Next, we rule out scenarios (*i*) and (*ii*) based on experimental data.

*(i) Ruling out a first-order transition* – (1) In Extended Data Fig. 8a & b, we plot data of both resistance $R_{xx}$ and Nernst signal $V_N$ as a function of density $n_g$, scanning $n_g$ in both directions. All first-order transitions lead to hysteresis in observables. To the highest experimental resolution in $R_{xx}$ and $V_N$, no hysteresis is observable. (2) In the phase diagram of Fig. 4a, we have experimentally demonstrated that the $n_g$-induced transition is precisely the zero-temperature limit of the temperature-driven Berezinskii–Kosterlitz–Thouless (BKT) transition, which is a continuous transition. (3) In Extended Data Fig. 8c & d, we carefully monitor how $R_{xx}$ deviates from zero as we cross the BKT boundary starting from the superconducting side, by either tuning $n_g$ or $T$. The curve of $R_{xx}(T)$ reflects the standard continuous transition (BKT). We find that the two curves $R_{xx}$ ($n_g$) and $R_{xx}(T)$ overlap excellently when the $n_g$ and $T$ axes are linearly scaled properly. This is strong evidence that the transition out of the superconducting state driven by decreasing $n_g$ at base $T$ is continuous just like the standard BKT transition (unbinding of vortex-antivortex pairs). (4) In our experiments, the vortex Nernst signal directly probes the vortices, further confirming a BKT-like transition approached from the superconducting side. All these experimental data support that the $n_g$-tuned superconducting QPT is continuous, which is natural and by default anyway for a 2D superconducting transition. Note that the vortex Nernst signal vanishes once $n_g$ drops below $n_c$, which is totally unexpected in the conventional BKT picture (vorticity enhances further the long fluctuation tail in the disordered normal state). This is the sudden death syndrome on the non-superconducting side that defines the central puzzle in our report.

(*ii*) *Ruling out two independent, but accidentally degenerate, continuous transitions* – This scenario involves two independent critical densities, one for the transition from the superconducting state (ordered) to a non-superconducting state (disordered normal state) at $n_{c1}$ and the other for the transition from this disordered non-superconducting state to another non-superconducting state (anomalous, ordered) at $n_{c2}$. (1) In our observations, the $n_g$-tuned QCP at zero $B$ is very sharp (Fig. 3a), i.e., the Nernst signal is sharply concentrated at a *single* critical density ($n_c$). It would be very unnatural to assume that within such a narrow regime, two independent critical densities can occur at the exact same density (i.e., it requires $n_{c1} = n_{c2} = n_c$). (2) Above the superconducting transition temperature (~ 900 mK), the conductivity near $n_c$ follows well with a linear relation to $n_g$ (Extended Data Fig. 5d); namely no evidence of any additional non-superconducting QPT (i.e., the transition at $n_{c2}$) is observed. (3) Importantly, in Extended Data Fig. 7, we demonstrate that a different device yields again a sharp, single critical density at which the QPT occurs, yet it is at a very different value (due to different impurity levels). If there were two independent densities that are accidentally degenerate in one device, they should not move together precisely in different devices. The only feasible explanation is that there is only one single critical density that connects the superconducting state ($n_g > n_c$) to a metallic-like non-superconducting state ($n_g < n_c$, regime III).

Nevertheless, there are also challenges in employing the concept of DQCP here. (1) The existing theoretical proposal of a DQCP at the QSHI-superconductor transition[14], albeit attractive, cannot be directly applied to our experimental case. The toy model assumes that the non-superconducting side (i.e., QSHI) at low temperature is an insulator. Yet, in our case, we observed a metallic-like (resistance ~ kΩ at millikelvins, with weak $T$-dependence) non-superconducting state below $n_c$. It hence challenges the theoretical modeling. An anomalous metal next to a superconducting state, even for a conventional QCP, is a challenging topic[2]. We don't know how this state may alter the picture of a DQCP. (2) In the proposed DQCP on a QSHI-superconductor transition[14], the QSHI order develops from spontaneous symmetry breaking. At the current stage, we don't know how this may happen in monolayer $WTe_2$. However, we note that it may be crucially important to consider the excitonic insulator state of monolayer $WTe_2$ in addressing our results. (3) Are there charged-2e skyrmions in regime III? Our experiments directly encourage the search for their presence in monolayer $WTe_2$. Their presence/absence is key to unlocking the secret. (4) The concept of DQCP is highly theoretical so far. Currently, a conclusive experimental demonstration of a DQCP remains absent, and there exists no established standard for arguing which types of experimental data can provide proof of this novel concept. In this regard, our results suggest that careful measurements of quantum fluctuations near the QCP can provide valuable guidance to the theory, and identify a realistic materials system (monolayer $WTe_2$) for testing various ideas. A comprehensive understanding of the observed phenomena here and a clarification of its relations to the DQCP require substantial inputs from both theory and future experiments.

**Extended Data Fig. 1**

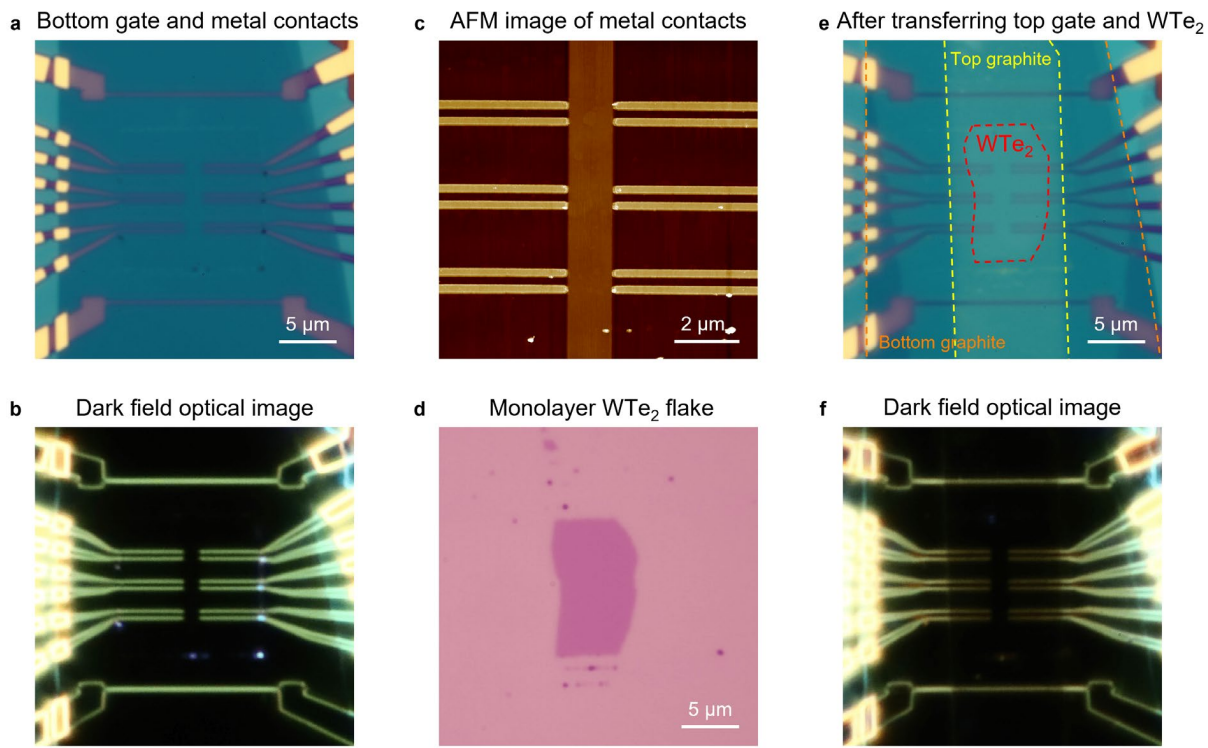

**Extended Data Fig. 1 | Microscope and AFM images during the device 1 fabrication process. a**, Bottom gate and metal electrodes after AFM tip clean. **b**, Dark field optical image of the bottom part. **c**, Tapping-mode AFM image of the metal electrodes. **d**, Microscope image of monolayer $WTe_2$ flake exfoliated in the glovebox. **e**, Microscope image of device 1 after transferring the top gate and monolayer $WTe_2$. The monolayer $WTe_2$ flake is outlined in red. Top and bottom graphite flakes are outlined in yellow and orange. **f**, Dark field optical image of device 1. No bubbles were observed.

**Extended Data Fig. 2**

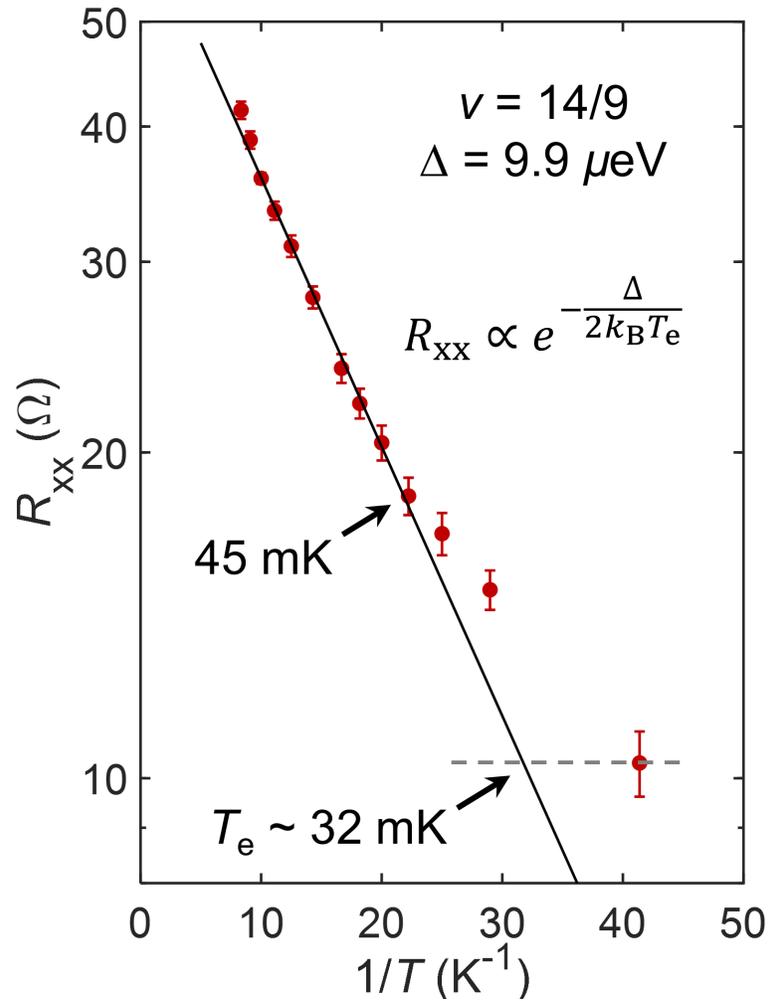

**Extended Data Fig. 2 | Electron temperature calibration of the measurements in the dilution refrigerator.** The calibration was performed based on an ultra-high mobility GaAs device with a series of fractional quantum Hall states. The plot shows the longitudinal resistance ($R_{xx}$) in log scale versus the inverse fridge temperature ($1/T$), taken at the 14/9 fractional quantum Hall state.

**Extended Data Fig. 3**

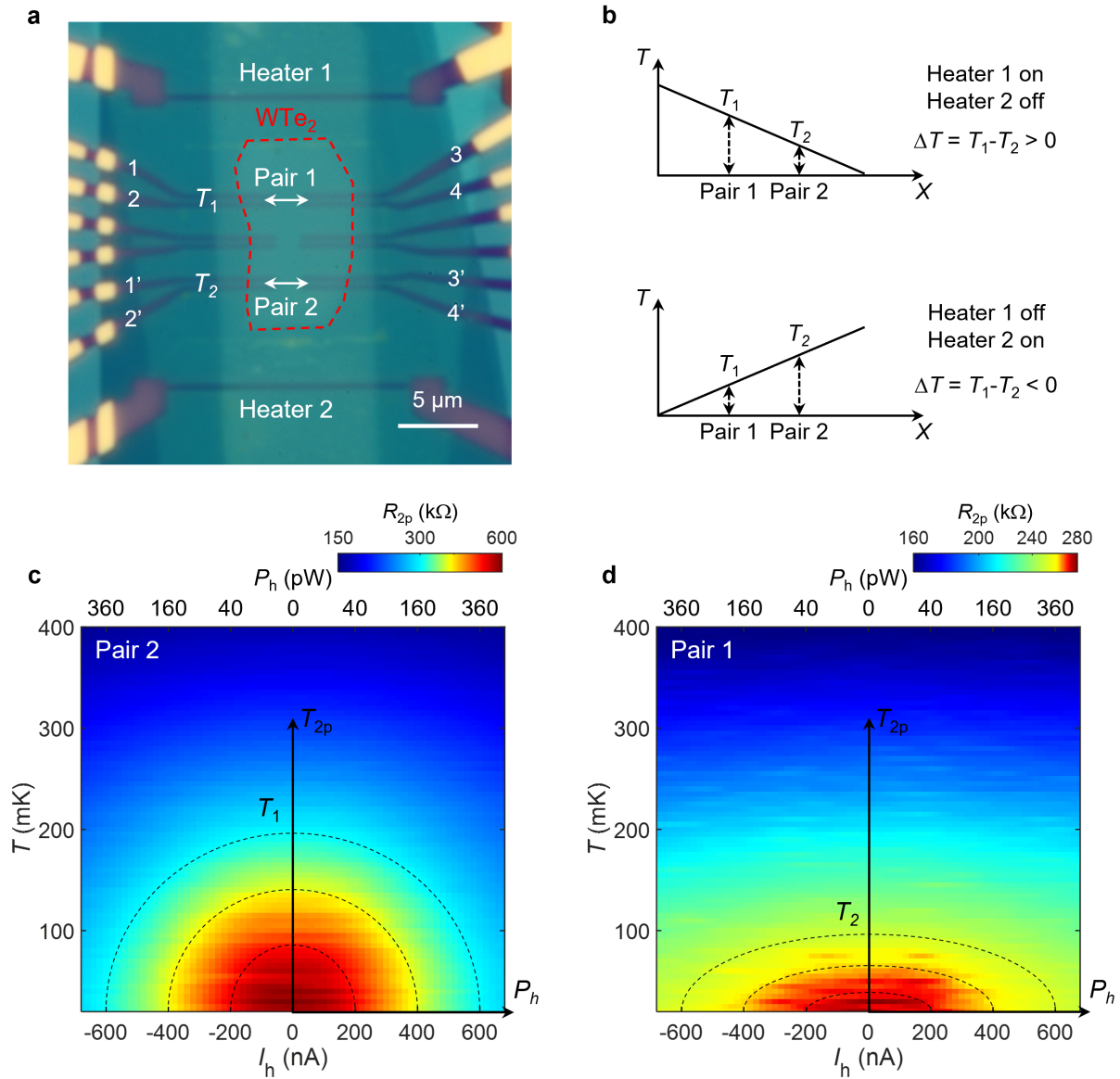

**Extended Data Fig. 3 | Temperature gradient estimation in device 1. a**, Microscope image of device 1 with the two-probe resistance measurement configuration. The monolayer $WTe_2$ flake is outlined in red. **b**, Temperature gradient generated by Heater 1 and Heater 2. **c** and **d**, Two-probe resistance for Pair 2 and Pair 1, respectively, as a function of the Heater 2 current and the fridge temperature. The black dashed lines trace the same $R_{2p}$, and extract $T_1$ and $T_2$.

**Extended Data Fig. 4**

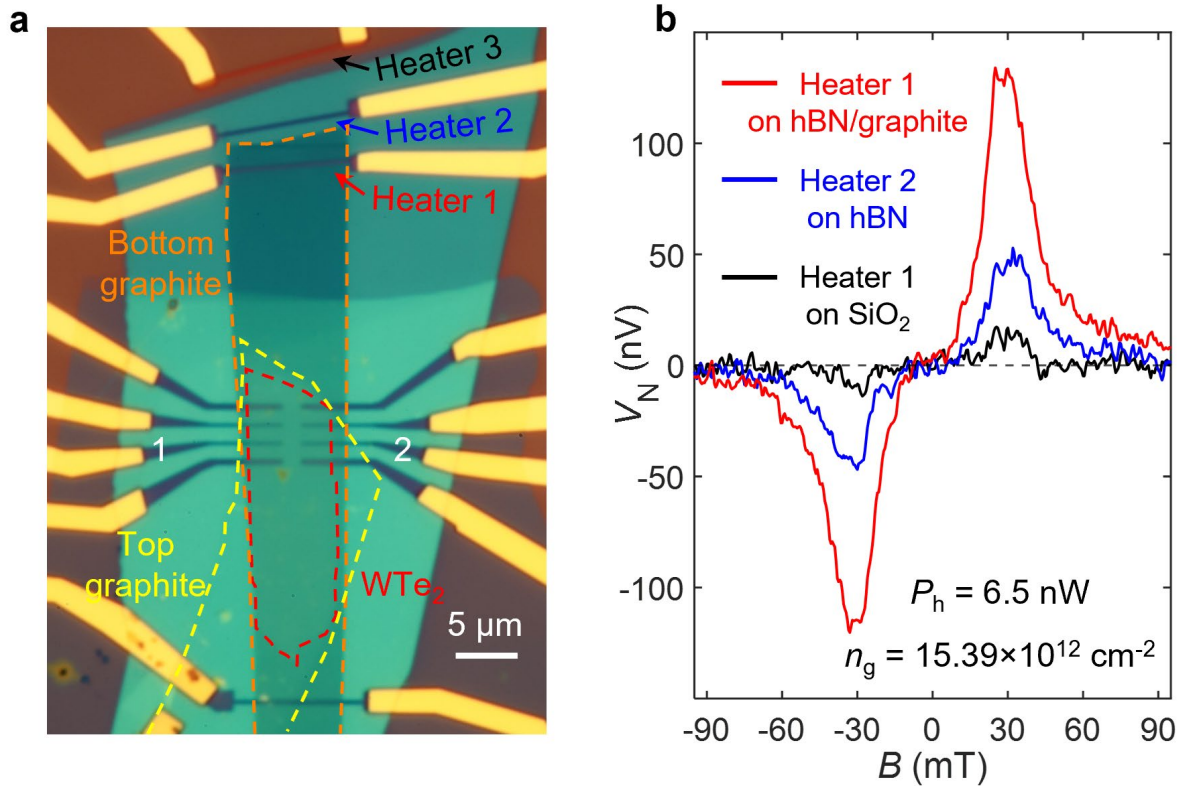

**Extended Data Fig. 4 | High efficiency of microheaters fabricated on vdW heterostructures. a**, Microscope image of device 2 with three microheaters fabricated on hBN/graphite (Heater 1), only hBN (Heater 2), and only SiO$_2$ (Heater 3), respectively. The monolayer WTe$_2$ flake is outlined in red. Top and bottom graphite flakes are outlined in yellow and orange. The Nernst signal was measured from the same two probes labeled 1 and 2. **b**, Nernst signal as a function of $B$ measured with Heater 1 (red), Heater 2 (blue), and Heater 3 (black), respectively, for the same $P_h$ = 6.5 nW and $n_g$ = 15.39×10$^{12}$ cm$^{-2}$.

**Extended Data Fig. 5**

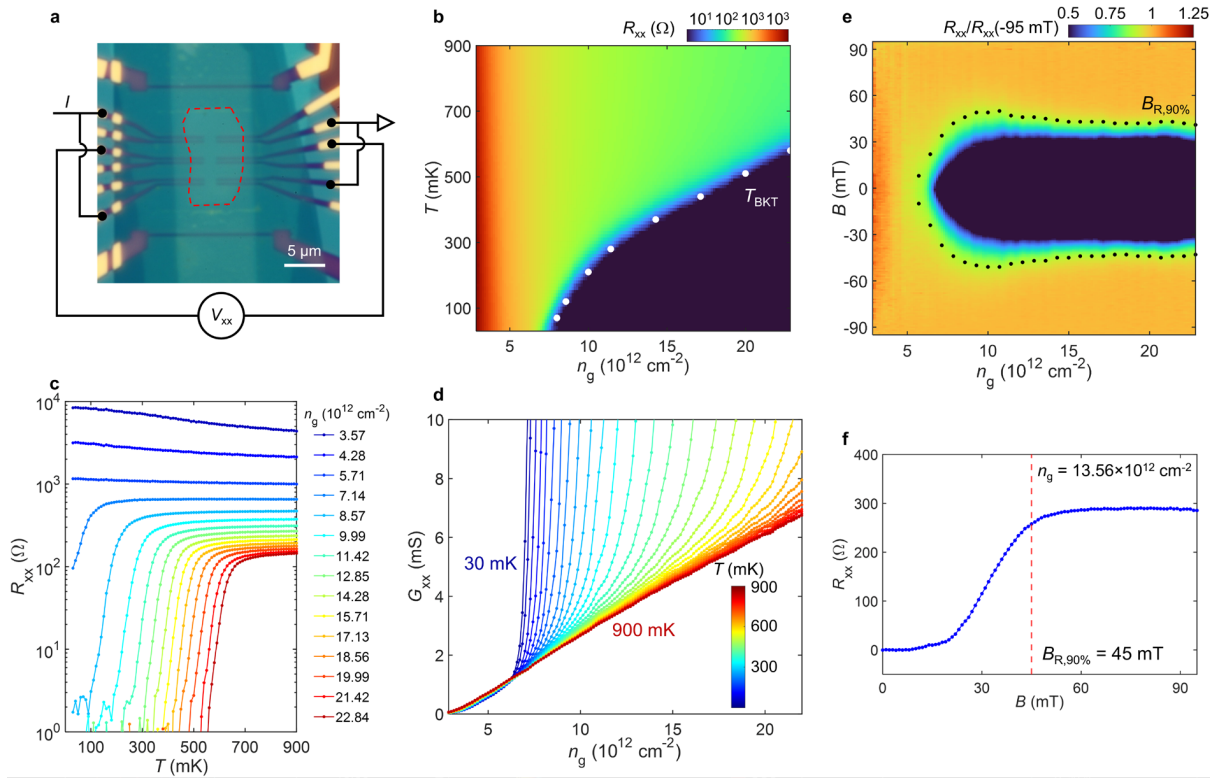

**Extended Data Fig. 5 | Raw resistance data taken on device 1. a**, Microscope image of device 1 with the four-probe resistance measurement configuration. The monolayer WTe$_2$ flake is outlined in red. **b**, Four-probe resistance as a function of $n_g$ and temperature. The white dots represent $T_{BKT}$. **c**, Temperature dependence of $R_{xx}$ for various $n_g$. **d**, The four-probe conductance ($G_{xx}$) calculated from $R_{xx}$ as a function of $n_g$ at various temperatures. **e**, Normalized $R_{xx}$ as a function of $n_g$ and $B$ (the same data of Fig. 2b). For each $n_g$, $R_{xx}$ is normalized to its value at -95 mT to highlight its $B$ dependence. The black dotted line represents $B_{R,90\%}$, at which $R_{xx}$ drops to 90% of its saturated value at high $B$. **f**, $R_{xx}$ as a function of $B$ for $n_g = 13.56 \times 10^{12}$ cm$^{-2}$ showing a vertical line cut of **e**. The red dashed line indicates the identified $B_{R,90\%}$.

**Extended Data Fig. 6**

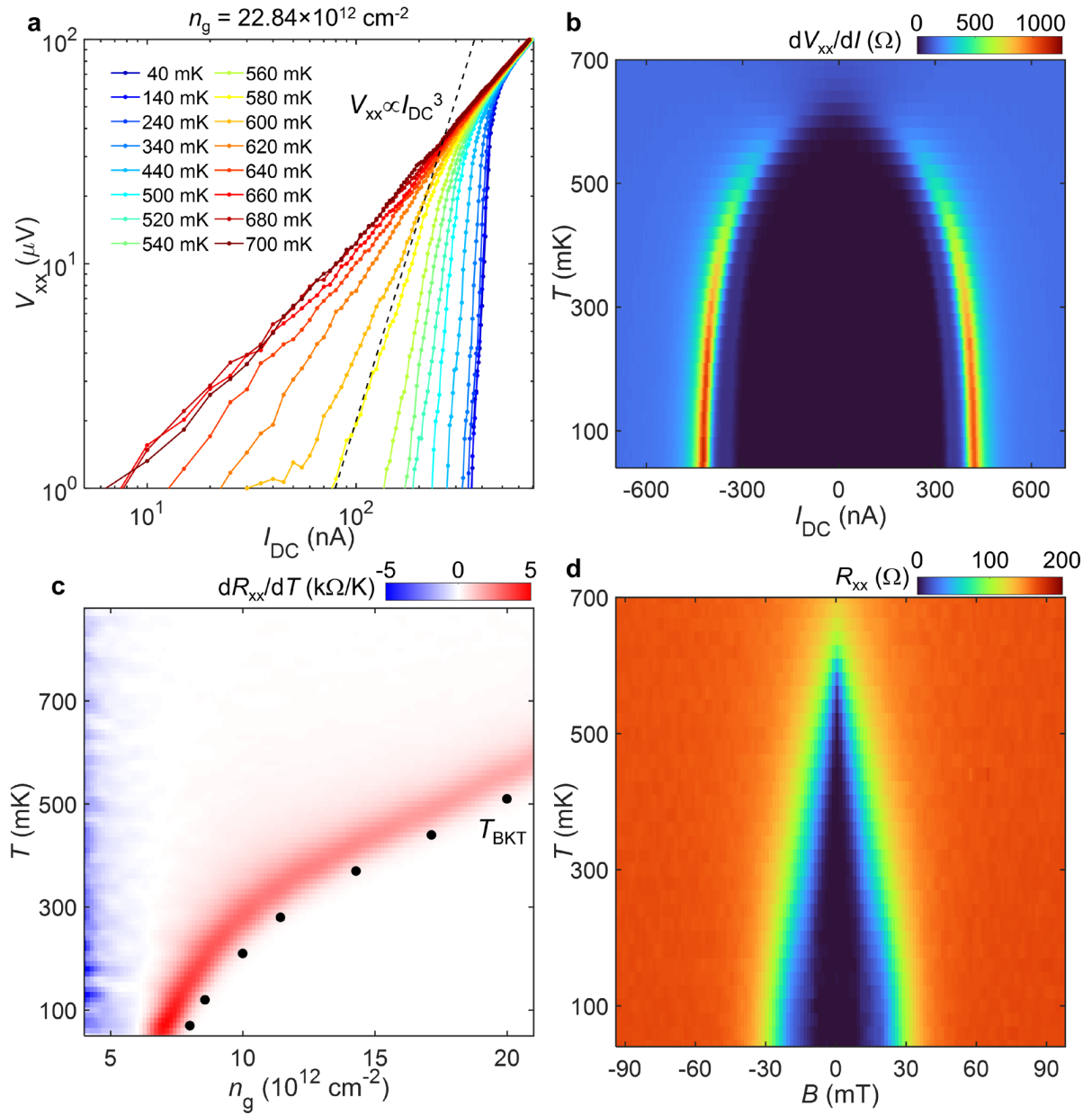

**Extended Data Fig. 6 | Extracting $T_{BKT}$ in device 1. a**, Characteristic nonlinear *I-V* curves on a logarithmic scale at various temperatures. The black dashed line shows a tentative fit to the $V_{xx} \propto I_{DC}^3$ power law, which extracts $T_{BKT}$ to be ~ 580 mK for $n_g = 22.84 \times 10^{12}$ cm$^{-2}$. **b**, The corresponding differential resistance ($dV_{xx}/dI$) as a function of $I_{DC}$ and *T*. **c**, $dR_{xx}/dT$ as a function of $n_g$ and *T* (the same data of Fig. 1b, using two neighboring points for the temperature derivative). The black dots represent $T_{BKT}$ extracted by fitting to the $V_{xx} \propto I_{DC}^3$ power law for various $n_g$. **d**, *B*-dependence of $R_{xx}$ as a function of *T* for $n_g = 22.84 \times 10^{12}$ cm$^{-2}$.

**Extended Data Fig. 7**

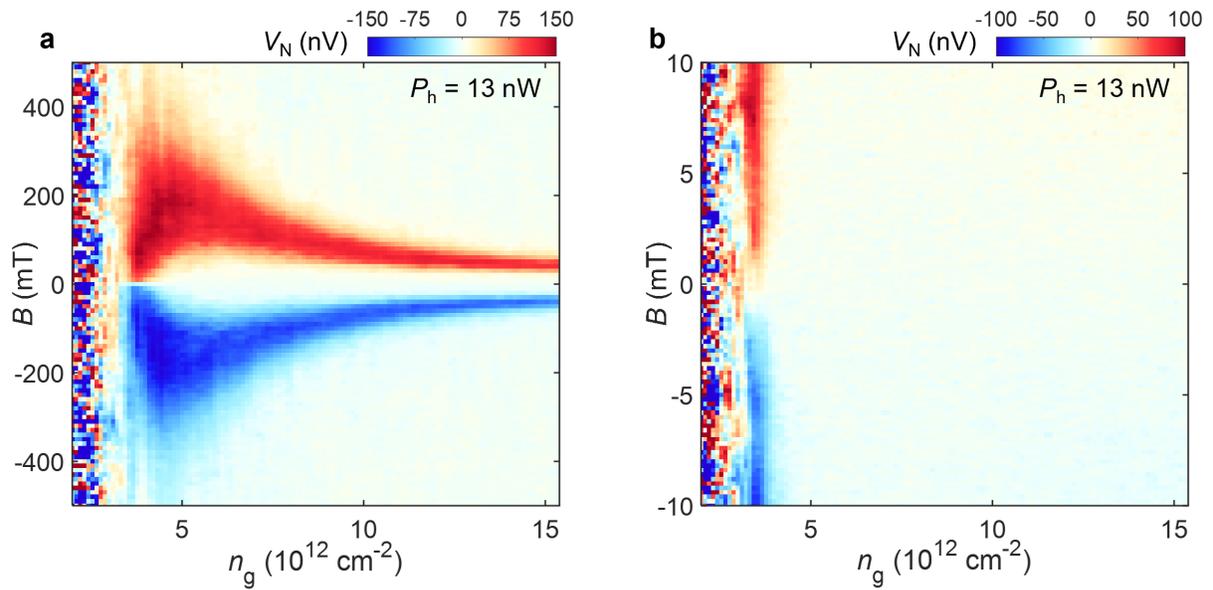

**Extended Data Fig. 7 | Nernst data taken on device 2. a**, Nernst signal as a function of $n_g$ and $B$. **b**, Nernst signal as a function of $n_g$ under very small $B$. The dual-heater power is 13 nW. The Nernst signal is recorded at the base temperature. The noise data (also seen in device 1) at low $n_g$ is not antisymmetric to $B$ and hence not Nernst signal. They likely arise due to bad contact to Au in this regime. The results reproduce the findings in device 1 discussed in the main text.

**Extended Data Fig. 8**

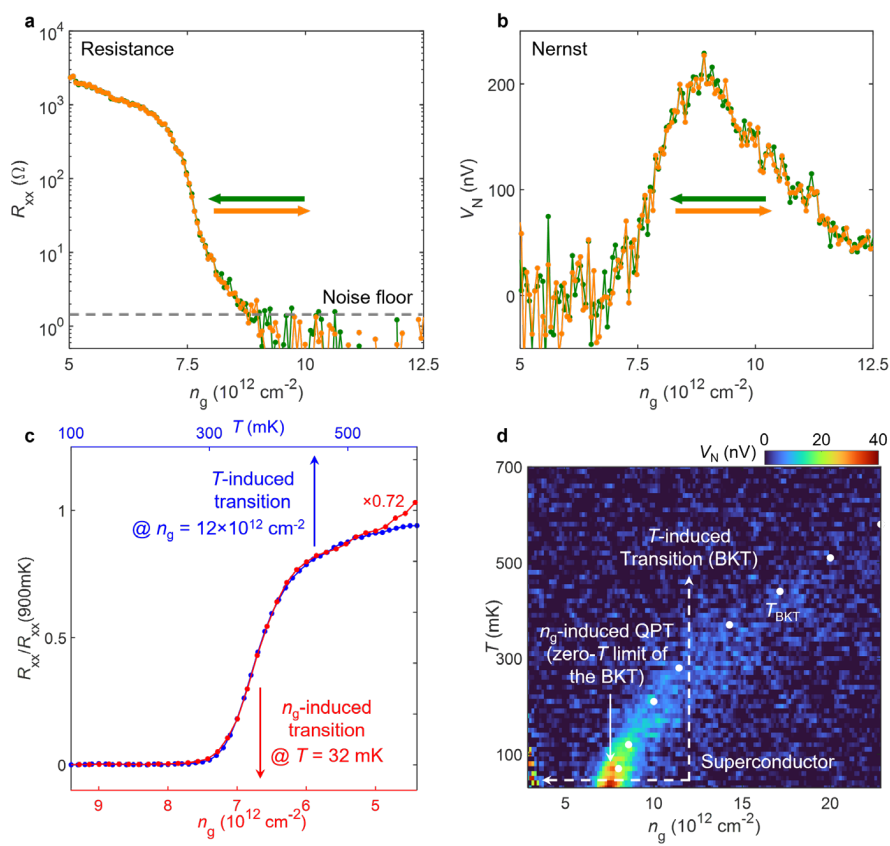

**Extended Data Fig. 8 | Ruling out a first-order transition. a**, $R_{xx}$ as a function of $n_g$, swept back and forth, at base temperature. The green and orange curves correspond to decreasing and increasing $n_g$, respectively. **b**, Nernst signal as a function of $n_g$ measured at 5 mT. The dual-heater power is 128 pW. No hysteresis in the sweeps, the characteristic of a first-order transition, is found for both resistance and Nernst signal. Note that data in **a** and **b** were taken from a new cooldown after about a year (stored in an Ar-filled glovebox) after the first cooldown, in which we took most of the data presented in the manuscript. The device and data remain of high quality with negligible changes. **c**, $R_{xx}$ normalized to its value at 900 mK as a function of $n_g$ (red) or $T$ (blue), along the two dashed lines shown in **d**. The data is extracted from Fig. 1b. $R_{xx}(n_g)$ and $R_{xx}(T)$ display an excellent overlap under linear scaling which suggests that the two transitions share the same characteristic. Started from the same superconducting state, the $T$-induced transition is a continuous BKT transition in nature, whereas the $n_g$-induced transition is also induced by the proliferation of vortices and antivortices (driven by quantum fluctuations). **d**, Nernst signal as a function of $n_g$ and $T$, taken at $B = 2$ mT and $P_h = 46$ pW (the same data from Fig. 4a). The $n_g$-induced QPT manifests as the zero-temperature limit of the continuous BKT transition. All these observations speak against a first-order transition and demonstrate that indeed the $n_g$-induced 2D superconducting transition is BKT-like, in terms of how the superconducting state is destroyed (the proliferation of vortices and antivortices). The sudden death of the Nernst signal below the critical density (Figs. 2 & 3) is unexpected, raising intriguing questions regarding the behaviors of the vortices near the QPT.